\documentstyle[12pt]{article}

\newcommand {\e} {\mbox{\rm e}}

\newcommand {\nn}    {\nonumber}
\newcommand {\vs}[1]  { \vspace*{#1 cm} }

\newcounter{eq}
\newcounter{sc}

\newcommand {\IJMP}  {Int. J. Mod. Phys.}
\newcommand {\MPL}  {Mod. Phys. Lett.}
\newcommand {\NP}   {Nucl. Phys.}
\newcommand {\PL}   {Phys. Lett.}

\newcommand {\AP}   {Ann. of Phys.}
\newcommand {\PTP}  {Prog. Theor. Phys.}

\newcommand {\JMP}  {J. Math. Phys.}

\def\overleftrightarrow#1{\vbox{\ialign{##\crcr
 $\leftrightarrow$\crcr\noalign{\kern-1pt\nointerlineskip}
 $\hfil\displaystyle{#1}\hfil$\crcr}}}

\newlength{\minitwocolumn}
\begin{document}

\begin{flushright}
EDO-EP-41\\
April, 2001\\
\end{flushright}
\vspace{30pt}

\pagestyle{empty}
\baselineskip15pt

\begin{center}
{\large\bf Higgs Mechanism with a Topological Term

 \vskip 1mm
}

\vspace{20mm}

Ichiro Oda
          \footnote{
          E-mail address:\ ioda@edogawa-u.ac.jp
                  }
\\
\vspace{10mm}
          Edogawa University,
          474 Komaki, Nagareyama City, Chiba 270-0198, JAPAN \\

\end{center}


\vspace{15mm}
\begin{abstract}
In cases of both abelian and nonabelian gauge groups, we consider 
the Higgs mechanism in topologically massive gauge theories in an arbitrary 
space-time dimension. It is shown that the presence of a topological term 
makes it possible to shift mass of gauge fields in a nontrivial way compared 
to the conventional value at the classical tree level.
We correct the previous misleading statement with respect to the counting
of physical degrees of freedom, where it is shown that gauge fields become
massive by 'eating' the Nambu-Goldstone boson and a higher-rank tensor field,
but a new massless scalar appears in the spectrum so the number of
the physical degrees of freedom remains unchanged before and after
the spontaneous symmetry breakdown. Some related phenomenological
implications 
and applications to superstring theory are briefly commented.
\vspace{15mm}

\end{abstract}

\newpage
\pagestyle{plain}
\pagenumbering{arabic}


\rm
\section{Introduction}

Despite its experimental success, the Standard Model based on
$SU(3)_C \times SU(2)_L \times U(1)_Y$ still leaves many questions
unsolved. For instance, why is a particular gauge group observed
at low energy, together with the multiplicity of generations?
Can we explain the origin of the proliferation of parameters in the
flavor sector from a fundamental theory? How do we incorporate
gravity into the theory?  It seems that the resolution of these problems
requires new conceptional framework and tools of the underlying
fundamental theory, that is, superstring theory.

In addition to these theoretical questions, the Higgs sector in the
Standard Model has not been observed experimentally and the least
understood.   
It is then of theoretical interest to ask what modification superstring
theory provides for the Higgs sector. However, superstring phenomenology,
the study of how superstring theory makes contact with physics at accessible 
energy, is still in its infancy, so we have no quantitative predictions, as 
yet, from superstring theory. Nevertheless, there are a number of important
qualitative implications and insights. In particular, superstring theory
predicts the existence of many new particles such as a dilaton, an axion
and perhaps other scalar moduli. In them, a bunch of antisymmetric tensor
fields also naturally appear in the spectrum and play an important role in
the 
non-perturbative regime, such as D-branes and various dualities, in
superstring theory \cite{Pol}. Thus it is natural to inquire if such 
antisymmetric tensor fields yield a new phenomenon to the still mysterious 
Higgs sector in the Standard Model.
 
In fact, it has been known that when there is a topological term,
antisymmetric 
tensor fields (including gauge field) exhibit an ingenious mass generation 
mechanism, which we call, the 'topological Higgs mechanism' in the sense that 
antisymmetric tensor fields acquire masses and spins without breaking the
local
gauge invariance explicitly. (Of course, in this case, unlike the
conventional 
Higgs mechanism, we do not have the Higgs particle in general, though.)
This interesting mass generation mechanism is first found within the framework
of three-dimensional gauge theory with Chern-Simons term \cite{Schonfeld,
Deser}.
Afterwards, this three-dimensional topological Higgs mechanism 
has been generalized to an arbitrary higher-dimension 
in cases of both abelian \cite{Minahan, Oda1, Allen}
and non-abelian gauge theories \cite{Oda2}. More recently, a new type of 
the topological massive nonabelian gauge theories with the usual Yang-Mills
kinetic term has been constructed \cite{Hwang, Landim, Harikumar}.

In this paper, we would like to study the mass generation mechanism in 
the abelian gauge theories \cite{Minahan, Oda1, Allen} and the nonabelian
gauge theories \cite{Hwang, Landim, Harikumar} with the usual Higgs
potential in addition to a topological term. Since in the Weinberg-Salam
theory the Higgs doublets plus their Yukawa couplings are one of the
key ingredients with great experimental success, we do not want to dismiss 
the framework of the conventional Higgs mechanism. Instead, we wish to
consider how mass of gauge fields is modified if the conventional Higgs
mechanism coexists with the topological Higgs mechanism. About ten years
ago, Yahikozawa and the present author have studied such a model, but
it is a pity that there is some misleading statement about the counting
of the degrees of freedom and only abelian gauge theories are taken into
consideration \cite{Oda3}.
 
The purpose of the present paper is threefold. First of all, we wish to
correct our previous misleading statement with respect to the counting
of physical degrees of freedom.  We show that gauge fields become
massive by 'eating' both the Nambu-Goldstone boson and a higher-rank 
tensor field,
but a new massless scalar appears in the spectrum so the number of
the physical degrees of freedom remains unchanged before and after
the spontaneous symmetry breakdown. Secondly, this phenomenon is generalized
to the nonabelian gauge theories. Finally, we wish to comment on some 
phenomenological implications and applications to superstring theory
of our results.

\rm
\section{Abelian gauge theories}

Let us start by reviewing the 'topological Higgs mechanism' in abelian 
gauge theories in an arbitrary space-time dimension \cite{Minahan}:
\begin{eqnarray}
S = - \frac{1}{2} \int dA \wedge \ast dA - \frac{1}{2} \int dB 
\wedge \ast dB + \mu \int A \wedge dB,
\label{1}
\end{eqnarray}
where $A$ and $B$ are respectively an $n$-form and a ($D-n-1$)-form
in $D$-dimensional space-time with metric signature $(-,+,+, \cdots,+)$,
"$\ast$" is the Hodge dual operator, and $\wedge$ is the Cartan's
wedge product, which we will omit henceforth for simplicity.
The equations of motion are easily derived to
\begin{eqnarray}
d \ast d A + (-1)^n \mu d B &=& 0, \nn\\
d \ast d B + (-1)^{n(D-1)+1} \mu d A &=& 0.
\label{2}
\end{eqnarray}
{}From these equations of motion, we obtain
\begin{eqnarray}
( \Delta + \mu^2 ) d A &=& 0, \nn\\
( \Delta + \mu^2 ) d B &=& 0,
\label{3}
\end{eqnarray}
where $\Delta$ is the Laplace-Beltrami operator.
Note that these equations are equations of motion for the transverse
components of $A$ and $B$ fields, and they clearly represent that
two massless antisymmetric tensor fields $A$ and $B$ have gained
the same mass $\mu$ through the topological term $\mu \int A \wedge dB$,
whose phenomenon we call the "topological Higgs mechanism".
It is worthwhile to count the degrees of freedom of physical states
before and after the topological Higgs mechanism occurs. Originally,
we have two massless fields, so the total number of the degrees of freedom
is given by ${D-2 \choose n} + {D-2 \choose D-n-1}$, which is rewritten
as 
\begin{eqnarray}
{D-2 \choose n} + {D-2 \choose D-n-1} &=& {D-1 \choose n} \nn\\
&=& {D-1 \choose D-n-1}.
\label{4}
\end{eqnarray}
This equation clearly indicates that the massless $n$-form field $A$ 
has become massive by 'eating' the massless ($D-n-1$)-form field $B$, 
and vice versa.

Now we would like to introduce the conventional Higgs potential for 
only $A$ field, for which we have to restrict $A$ field to be
a 1-form since the exterior derivative couples with only 1-form
to make the covariant derivative. Thus we are led to consider 
\cite{Oda3}
\begin{eqnarray}
S &=& - \frac{1}{2} \int dA \ast dA - \frac{1}{2} \int dB \ast dB 
+ \mu \int A dB  \nn\\
&-& \int \Big[ (D \phi)^\dagger \ast D \phi 
+ \lambda (|\phi|^2 - \frac{1}{2} v^2) \ast (|\phi|^2 - \frac{1}{2} 
v^2) \Big],
\label{5}
\end{eqnarray}
where $B$ is now a ($D-2$)-form field, and $\phi$ and $v$ are respectively a
0-form complex field and a real number. The covariant derivative $D$ is
defined in a usual way as
\begin{eqnarray}
D \phi &=& d \phi - i g A \phi, \nn\\
(D \phi)^\dagger &=& d \phi^\dagger + i g A \phi^\dagger.
\label{6}
\end{eqnarray}
The gauge transformations are given by
\begin{eqnarray}
\phi(x) &\rightarrow& \phi'(x) = \e^{- i \alpha(x)} \phi(x), \nn\\
A(x) &\rightarrow& A'(x) = A(x) - \frac{1}{g} d \alpha(x), \nn\\
B(x) &\rightarrow& B'(x) = B(x) - d \beta(x),
\label{7}
\end{eqnarray}
where $\alpha(x)$ and $\beta(x)$ are a 0-form and a ($D-3$)-form
gauge parameters, respectively. And $g$ denotes a $U(1)$ gauge
coupling constant. Note that there are still off-shell reducible
symmetries for $B$ field when $D > 3$. 

The minimum of the potential is achieved at 
\begin{eqnarray}
|\phi| = \frac{v}{\sqrt{2}},
\label{8}
\end{eqnarray}
which means that the field operator $\phi$ develops a vacuum
expectation value $|<\phi>| = \frac{v}{\sqrt{2}}$. If we write $\phi$
in terms of two real scalar fields $\phi_1$ and $\phi_2$ as
\begin{eqnarray}
\phi = \frac{1}{\sqrt{2}} ( \phi_1 + i \phi_2 ),
\label{9}
\end{eqnarray}
we can select 
\begin{eqnarray}
<\phi_1> = v,  \ <\phi_2> = 0.
\label{10}
\end{eqnarray}
With the shifted fields
\begin{eqnarray}
\phi'_1 = \phi_1 - v,  \  \phi'_2 = \phi_2,
\label{11}
\end{eqnarray}
we have 
\begin{eqnarray}
<\phi'_1> = <\phi'_2> = 0.
\label{12}
\end{eqnarray}
Note that $\phi'_2$ corresponds to the massless Goldstone
boson.  At this stage, let us take the unitary gauge to remove
the mixing term between $A$ and $\phi'_2$ in the action:
\begin{eqnarray}
\phi^u(x) &=& \e^{-i \frac{1}{v} \xi(x)} \phi(x) = \frac{1}{\sqrt{2}} 
( v + \eta(x) ), \nn\\
G_\mu(x) &=&  A_\mu(x) - \frac{1}{gv} \partial_\mu \xi(x).
\label{13}
\end{eqnarray}
In this gauge condition, $\xi(x)$ and $\eta(x)$ correspond to
$\phi'_2(x)$ and $\phi'_1(x)$, respectively. Also note that the
unitary gauge corresponds to the gauge transformation with 
a fixed gauge parameter $\alpha(x) = \frac{1}{v} \xi(x)$. Then, the 
action (\ref{5}) reduces to the form
\begin{eqnarray}
S &=& - \frac{1}{2} \int dG \ast dG - \frac{1}{2} (g v)^2 \int G \ast G
- \frac{1}{2} \int dB \ast dB \nn\\
&+& \mu \int G dB - \int \Big[ \frac{1}{2} d\eta \ast d\eta + \frac{1}{2} 
(\sqrt{2 \lambda} v)^2 \eta \ast \eta \Big] \nn\\
&-& \int \Big[ \frac{1}{2} g^2 \eta (2 v + \eta) G \ast G 
+ \lambda (v \eta \ast \eta^2 + \frac{1}{4} \eta^2 \ast \eta^2) \Big].
\label{14}
\end{eqnarray}
{}From the above action, we can easily read off that the Higgs field 
$\eta(x)$ becomes a massive field with mass $\sqrt{2 \lambda} v$ and 
the would-be Goldstone boson $\xi(x)$ is absorbed into the gauge field 
$G_\mu$, as in the conventional Higgs mechanism without a topological 
term.

In order to clarify the mass generation mechanism for the gauge
field $G$ and the antisymmetric tensor field $B$, it is enough to
consider only the quadratic terms with respect to fields in the
action (\ref{14}). In other words, neglecting the interaction terms
in the action (\ref{14}), we  derive the equations of motion for $G$ 
and $B$ fields whose concrete expressions are given by
\begin{eqnarray}
- d \ast dG - (gv)^2 \ast G + \mu dB &=& 0. \\
(-)^D d \ast dB + \mu dG &=& 0.
\label{15,16}
\end{eqnarray}
{}From these equations, we can obtain 
\begin{eqnarray}
\Big[ \Delta + \mu^2 + (gv)^2 \Big] dG &=& 0. \\
\Big[ \Delta + \mu^2 + (gv)^2 \Big] \delta dB &=& 0. \\
\delta G &=& 0,
\label{17,18,19}
\end{eqnarray}
where $\delta$ denotes the adjoint operator. Eqs. (17) and (18) reveal
that fields $G$ and $B$ have become massive fields with the same mass 
$\sqrt{\mu^2 + (gv)^2}$ through the conventional and topological Higgs
mechanisms. Also note that since Eq. (19) holds only when $g v \ne 0$, 
the existence of Eq. (19) reflects a characteristic feature in the case 
at hand.

Now let us consider how the conventional and topological Higgs mechanisms
have worked. For this, it is useful to count physical degrees of freedom
in the present model. Before spontaneous symmetry breaking, we have two
real scalar fields $\phi_1$ and $\phi_2$, and two massless fields $A_\mu$
and $B_{\mu_1 \cdots \mu_{D-2}}$. The total number of the degrees of freedom
is
\begin{eqnarray}
2 + {D-2 \choose 1} + {D-2 \choose D-2} = D + 1.
\label{20}
\end{eqnarray}
On the other hand, after the symmetry breaking, we have one real scalar
field and one massive field $G_\mu$ (or equivalently, $B_{\mu_1 \cdots 
\mu_{D-2}}$), so it appears that the total number of the degrees of 
freedom is now given by
\begin{eqnarray}
1 + {D-1 \choose 1} = 1 + {D-1 \choose D-2} = D.
\label{21}
\end{eqnarray}
However, we encounter a mismatch of one degree of freedom before and after
spontaneous symmetry breaking, which is precisely a question raised in 
our previous paper \cite{Oda3}. It is a pity that we have proposed
a misleading answer to this question in the previous paper, so
we wish to give a correct answer in this paper.

To find a missing one physical degree of freedom, we have to
return to the original equations of motion (15) and (16). From
Eq. (16), we have
\begin{eqnarray}
\ast dB + (-)^D \mu G = d\Lambda,
\label{22}
\end{eqnarray}
where $\Lambda$ is a 0-form. Then, using (19) we obtain
\begin{eqnarray}
\Delta \Lambda = 0,
\label{23}
\end{eqnarray}
which means that $\Lambda$ is a massless real scalar field that
we have sought. \footnote{Indeed, by performing the manifestly 
covariant quantization, we can show that this field is a physical
field with the positive norm \cite{Oda4}.}
By counting this one degree of freedom, we have
$D + 1$ degrees of freedom after the symmetry breaking which
coincides with the number of the degrees of freedom before the 
symmetry breaking.
In order to understand this degree of freedom further, using Eqs. (15),
(19) and (22), let us derive equations of motion for $G$ and $dB$.
The result reads
\begin{eqnarray}
\Big[ \Delta + \mu^2 + (gv)^2 \Big] G - (-1)^D \mu d\Lambda &=& 0, 
\nn\\
\Big[ \Delta + \mu^2 + (gv)^2 \Big] dB - (-1)^D (gv)^2 \delta \ast 
\Lambda &=& 0.
\label{24}
\end{eqnarray}
In Eq. (17), the last term involving $\Lambda$ in the former equation of
(\ref{24}) is projected out by the differential operator $d$. Similarly,
in Eq. (18), the last term in the latter equation of (\ref{24}) is also
projected out owing to the adjoint operator $\delta$.
As a final remark, it is worth emphasizing that $\Lambda$ is not the 
Goldstone boson $\xi$ but a new boson whose fact can be understood by
comparing (13) with (22).
Roughly speaking, $G_\mu$ (or $B_{\mu_1 \cdots \mu_{D-2}}$) has become 
massive by 'eating' both the Nambu-Goldstone boson $\xi$ and 
$B_{\mu_1 \cdots \mu_{D-2}}$ (or $G_\mu$), but it has eaten too much
more than its capacity! In consequence, a new scalar field has been 
vomitted.

\rm
\section{Nonabelian gauge theories}

We now turn to nonabelian theories. Let us start by reviewing the 
topologically massive nonagelian gauge theories \cite{Hwang, Landim, 
Harikumar}.
The action reads
\begin{eqnarray}
S = \int Tr \Big[ - \frac{1}{2} F \ast F - \frac{1}{2} {\cal{H}} 
\ast {\cal{H}} + \mu B F \Big],
\label{25}
\end{eqnarray}
where we use the following definitions and notations: $F = dA + g A^2, 
H = DB = dB + g [A, B], {\cal{H}} = H + g[F, V]$ and the square 
bracket denotes the graded bracket $[P, Q] = P \wedge Q - (-1)^{|P||Q|} 
Q \wedge P$. And $A$, $B$ and $V$ are respectively a 1-form, a ($D-2$)-form,
a ($D-3$)-form. All the fields are Lie group valued, for instance,
$A = A^a T^a$ where $T^a$ are the generators.
This action is invariant under the gauge transformations
\begin{eqnarray}
\delta A &=& D \theta = d\theta + g [A, \theta], \nn\\
\delta B &=& D \Omega + g [B, \theta], \nn\\
\delta V &=& - \Omega + g [V, \theta], 
\label{26}
\end{eqnarray}
where $\theta$ and $\Omega$ are a 0-form and a ($D-3$)-form gauge
parameters. Recall that a new field strength ${\cal{H}}$ together with
an auxiliary field $V$ has been introduced to compensate for the 
non-invariance of the usual kinetic term $Tr H^2$ under the tensor 
gauge transformations associated with $B$ \cite{Thierry}. From now on, 
we shall set a coupling constant $g$ to be 1 for simplicity 
since we can easily recover it whenever we want.

The equations of motion take the forms
\begin{eqnarray}
D \ast F &=& -[B, \ast {\cal{H}}] + D [V, \ast {\cal{H}}] + \mu DB, \nn\\
D \ast {\cal{H}} &=& (-1)^{D-1} \mu F, \nn\\
{}[F, \ast {\cal{H}}] &=& 0. 
\label{27}
\end{eqnarray}
{}From these equations, we can derive the following equations
\begin{eqnarray}
D \ast D \ast F + (-1)^D \mu^2 F &=& - \mu D \ast [F, V] - D \ast 
[B, \ast {\cal{H}}] + D \ast D [V, \ast {\cal{H}}], \nn\\
D \ast D \ast {\cal{H}} + (-1)^D \mu^2 {\cal{H}} &=& -(-1)^D \mu 
\Big( - \mu [F, V] - [B, \ast {\cal{H}}] + D [V, \ast {\cal{H}}]
\Big). 
\label{28}
\end{eqnarray}
A linear approximation for fields in (28) leads to equations
\begin{eqnarray}
( \Delta + \mu^2 ) dA &=& 0, \nn\\
( \Delta + \mu^2 ) dB &=& 0,
\label{29}
\end{eqnarray}
which imply that the fields $A$ and $B$ become massive by the 
topological Higgs mechanism as in the abelian gauge theories.

Next let us couple the Higgs potential to the model.
For simplicity and definiteness, we shall take the gauge group
to be $G = SU(2)$, with generators $T^i$ satisfying
\begin{eqnarray}
[ T^i, T^j] &=& i \varepsilon^{ijk} T^k, \nn\\
Tr(T^i T^j) &=& \frac{1}{2} \delta^{ij}, \nn\\
T^i &=& \frac{1}{2} \tau^i,
\label{30}
\end{eqnarray}
where $\tau^i$ are the Pauli matrices.
Then, the action is given by
\begin{eqnarray}
S &=& \int Tr \Big[ - F \ast F - {\cal{H}} \ast {\cal{H}} 
+ 2 \mu B F \Big] \nn\\
&-& \int \Big[ (D \phi)^\dagger \ast D \phi 
+ \lambda (|\phi|^2 - \frac{1}{2} v^2) \ast (|\phi|^2 - \frac{1}{2} 
v^2) \Big],
\label{31}
\end{eqnarray}
where $F^i_{\mu\nu} = \partial_\mu A^i_\nu - \partial_\nu A^i_\mu
+ \varepsilon^{ijk} A^j_\mu A^k_\nu$ and $D_\mu \phi = ( \partial_\mu
- i \frac{1}{2} \tau^i A^i_\mu) \phi$.

As in the abelian theories, let us take the unitary gauge given by
\begin{eqnarray}
\phi^u(x) &=& U(x) \phi(x) = {0 \choose \frac{1}{\sqrt{2}} 
( v + \eta(x) )}, \nn\\
G(x) &=&  U(x) A(x) U(x)^{-1} + i U(x) dU(x)^{-1}, \nn\\
B'(x) &=&  U(x) B(x) U(x)^{-1}, \nn\\
V'(x) &=&  U(x) V(x) U(x)^{-1},
\label{32}
\end{eqnarray}
where we have defined as $U(x) = \e^{-i \frac{1}{v} \tau^i \xi^i(x)}$.
It then turns out that the action (\ref{31}) reduces to the form
\begin{eqnarray}
S &=& \int Tr \Big[ - F \ast F - {\cal{H}} \ast {\cal{H}} 
+ 2 \mu B F \Big] \nn\\
&-& \int \Big[ (D \phi^u)^\dagger \ast D \phi^u + 
\lambda (v \eta + \frac{1}{2} \eta^2) \ast (v \eta + \frac{1}{2} \eta^2)
\Big],
\label{33}
\end{eqnarray}
where we have rewritten $B'$ and $V'$ as $B$ and $V$, respectively.
Namely, we now have the expressions like $F = dG + G^2$, $D \phi^u 
= d \phi^u - i G \phi^u$. In order to study the mass generation mechanism,
it is sufficient to examine only the quadratic action, which is given by
\begin{eqnarray}
S_0 &=& \int Tr \Big[ - dG \ast dG - dB \ast dB  + 2 \mu B dG \Big] \nn\\
&-& \int \Big[ \frac{1}{2} d\eta \ast d\eta + \frac{1}{2} 
\Big( \frac{gv}{2} \Big)^2
G^i \ast G^i + \lambda v^2 \eta \ast \eta \Big].
\label{34}
\end{eqnarray}
Here we have recovered the coupling constant $g$.
{}From this action (\ref{34}), it is easy to obtain the equations of motion 
\begin{eqnarray}
- d \ast dG^i - \Big( \frac{gv}{2} \Big)^2 \ast G^i + \mu dB^i &=& 0. \\
(-)^D d \ast dB^i + \mu dG^i &=& 0. \\
d \ast d\eta - (\sqrt{2 \lambda} v)^2 \ast \eta &=& 0.
\label{35,36,37}
\end{eqnarray}
{}From these equations, we can obtain 
\begin{eqnarray}
\Big[ \Delta + \mu^2 + \Big(\frac{gv}{2}\Big)^2 \Big] dG^i &=& 0. \\
\Big[ \Delta + \mu^2 + \Big(\frac{gv}{2}\Big)^2 \Big] \delta dB^i &=& 0. \\
\delta G^i &=& 0. \\
\Big[ \Delta + (\sqrt{2 \lambda} v)^2 \Big] \eta &=& 0.
\label{38,39,40,41}
\end{eqnarray}
Eq. (41) shows that the field $\eta$ is indeed the Higgs particle with mass 
$\sqrt{2 \lambda} v$ as in the conventional Higgs mechanism. Also note that 
Eqs. (38)-(40) are the same equations as Eqs. (17)-(19) except the $SU(2)$
index $i$ and the replacement $gv \rightarrow \frac{gv}{2}$, 
so we can show that a completely similar mass generation
mechanism to that in the abelian case occurs also in this
case. Thus the original $SU(2)$ gauge symmetry is completely broken and
all gauge fields (or antisymmetric tensor field) acquire the same mass 
$\sqrt{\mu^2 + (\frac{gv}{2})^2}$ via the conventional and topological Higgs
mechanisms. At the same time, we have a massless scalar with an $SU(2)$
index.

In this case as well, we can check the coincidence of the physical 
degrees of freedom before and after the symmetry breaking as follows.
Before the symmetry breakdown, 
\begin{eqnarray}
4 + 3 \times {D-2 \choose 1} + 3 \times {D-2 \choose D-2} = 3D + 1,
\label{42}
\end{eqnarray}
where 4 is the number of the physical degrees of freedom associated with
the complex Higgs doublet ($\phi$), and 3 appearing in the second and
third terms comes from $SU(2)$. On the other hand, after the symmetry
breakdown
\begin{eqnarray}
1 + 3 + 3 \times {D-1 \choose 1} = 3D + 1,
\label{43}
\end{eqnarray}
where 1 and 3 in the first and second terms equal to the numbers of 
the physical degrees of freedom of the real physical Higgs field
($\eta$) and the new massless field ($\Lambda^i$), respectively.

Finally, we wish to address a few topics related to the mechanism clarified 
in the present paper.
Firstly, we would like to consider how our model sheds some light on 
the Standard Model. For definiteness, let us consider the Weinberg-Salam
model on the basis of $SU(2)_L \times U(1)_Y$ even if it is easy
to extend the model at hand to the Standard Model based on $SU(3)_C \times
SU(2)_L \times U(1)_Y$. 
Note that we can construct a Weinberg-Salam model with topological terms
by unifying the abelian model treated in the previous section and
the $SU(2)$ nonabelian model in this section in an 
$SU(2)_L \times U(1)_Y$-invariant way. Then, we can observe the following
facts: In the conventional Weinberg-Salam model, mass of weak bosons is given 
by $M_W = \frac{gv}{2}$, whereas in our model it is given by 
$m_W = \sqrt{\mu^2 + (\frac{gv}{2})^2}$. Similarly, in our model mass of $Z$ 
boson receives a contribution from a topological term. Concerning the Higgs 
particle, we have the same mass $\sqrt{2 \lambda} v$ in both the models.
Fermion 
masses are also the same in both the models. Thus, we can conclude that
compared 
to the conventional Standard Model, in our model with topological terms we
can 
in general introduce additional parameters stemming from topological terms in 
the mass formulas of gauge bosons without violating the local gauge
symmetries 
explicitly and changing the overall structure of the Standard Model. 
In turn, provided that experiment would predict $\mu \approx 0$ in future 
it seems that we need to propose some mechanism to suppress the
contribution to 
mass of gauge bosons, since there is $\it{a \ priori}$ no local symmetry
prohibiting
the appearance of topological terms. Moreover, our present model predicts the 
existence of a new massless scalar, which is in a sharp contrast to the 
Weinberg-Salam model where there is no such a massless boson. These distinct
features in the model at hand will be testable by future experiments.

Secondly, as mentioned in the introduction, many of antisymmetric tensor
fields
naturally appear in the spectrum in superstring theory. For instance,
the action of a supersymmetric $D3$-brane in Type IIB superstring
theory includes topological terms among antisymmetric tensor fields in the 
Wess-Zumino term as well as their kinetic terms in the Born-Infeld action
\cite{Cederwall, Bergshoeff, Aganagic, Kimura1, Kimura2}. 
Hence, it is expected that our model would have some implications in the 
non-perturbative regime of superstring theory. We wish to stress again that 
antisymmetric tensor fields and their topological coupling play a critical
role 
in string dualities \cite{Pol}.

Thirdly, as another implication to superstring theory, note that we have
the term 
$\int_{M_{10}} B \wedge X_8$ in the effective action of superstring theory for
the Green-Schwarz anomaly cancellation. This term yields a topological term
upon compactification to four dimensions. If the Higgs potential appears 
in addition to the topological term via a suitable compactification, the mass 
generation mechanism discussed in the paper would work nicely \cite{Oda3}.

\rm
\section{Conclusion}

In this paper, we have investigated a physical situation where
a topological term coexists with the Higgs potential. We have
found that the gauge field becomes massive by 'eating' the Nambu-Goldstone
boson and the antisymmetric tensor field, and 'vomits'
a new massless scalar field. Moreover, we have pointed out that 
when our model is extended to a more realistic model such as the 
Weinberg-Salam model and the Standard Model, it gives us some distinct
results, those are, the shift of mass of gauge fields and the presence
of a new massless boson, which would be checked in future by experiment.
Although the experiment might preclude our model, it would be necessary 
to propose some mechanism for suppressing the effects coming from a
topological term since there is no symmetry prohibiting the existence
of such a topological term. This problem becomes more acute in 
superstring-inspired phenomenology since superstring theory gives
rise to many antisymmetric tensor fields with a topological term
at low energy.

To make the present model viable, we need to give the proof
of unitarity and renormalizability. We wish to attack this proof
within the framework of perturbation theory in future.

\vs 1



\begin{thebibliography}{99}

\bibitem{Pol}
J. Polchinski, {{\it String Theory}, Cambridge University
Press (1998).}

\bibitem{Schonfeld}
J.F. Schonfeld, {\NP \ {\bf B185} (1981) 157.}

\bibitem{Deser}
S. Deser, R. Jackiw and S. Templeton, {\AP \ {\bf 140} (1982) 372.}

\bibitem{Minahan}
J.A. Minahan and R.C. Warner, {Florida Univ. preprint, UFIFT-HEP-89-15.}

\bibitem{Oda1}
I. Oda and S. Yahikozawa, {\PTP \ {\bf 83} (1990) 991.}

\bibitem{Allen}
T.J. Allen, M.J. Bowick and A. Lahiri, {\MPL \ {\bf A6} (1991) 559.}

\bibitem{Oda2}
I. Oda and S. Yahikozawa, {\PL \ {\bf B234} (1990) 69.}

\bibitem{Hwang}
D.S. Hwang and C.-Y. Lee, {\JMP \ {\bf 38} (1997) 30, hep-th/9512216.}

\bibitem{Landim}
R.R. Landim and C.A. Almeida, {hep-th/0010050.}

\bibitem{Harikumar}
E. Harikumar, A. Lahiri and M. Sivakumar, {hep-th/0102013.}

\bibitem{Oda3}
I. Oda and S. Yahikozawa, {\MPL \ {\bf A6} (1991) 295.}

\bibitem{Oda4}
I. Oda, {to appear.}

\bibitem{Thierry}
J. Thierry-Mieg and L. Baulieu, {\NP \ {\bf B228} (1983) 259;
J. Thierry-Mieg and Y. Ne'eman, Proc. Natl. Acad. Sci. USA {\bf 79}
(1982) 7068.}

\bibitem{Cederwall}
M. Cederwall, A. von Gussich, B.E.W. Nilsson, P. Sundell
and A. Westerberg, {\NP \ {\bf B490} (1997) 163,
hep-th/9610148.}

\bibitem{Bergshoeff}
E. Bergshoeff and P.K. Townsend, {\NP \ {\bf B490} (1997) 145,
hep-th/9611173.}

\bibitem{Aganagic}
J. Aganagic, J. Park, C. Popescu, and J.H. Schwarz, {\NP \ {\bf B496}
(1997) 215, hep-th/9702133.}

\bibitem{Kimura1}
T. Kimura and I. Oda, {\NP \ {\bf B551} (1999) 183, hep-th/9811134.}

\bibitem{Kimura2}
T. Kimura and I. Oda, {\IJMP \ {\bf A16} (2001) 503, hep-th/9904019.}

\end{thebibliography}
\end{document}